\documentclass{aa}
\usepackage{amsmath} 
\usepackage{natbib}
\bibpunct{(}{)}{;}{a}{}{,} 
\usepackage[varg]{txfonts}
\usepackage{pgfplotstable}
\usepackage{tabularx}
\usepackage{float}
\usepackage{url}
\usepackage{multicol}
\usepackage{hyperref}

\usepackage{xcolor}

\usepackage{graphicx}
\usepackage{subfig}
\graphicspath{{./}{figures/}}
\pgfplotsset{compat=1.18}

\begin{document}

\title{Background exoplanet candidates in the original \textit{Kepler} field}

\author{J. Bienias \inst{1} \thanks{corresponding author: john.bienias@csfk.org}
\and
R. Szab\'o\inst{1,2}
}

\authorrunning{Bienias 
\and
Szab\'o
}

\titlerunning{Exoplanet Candidates in the Kepler Background}

\institute{Konkoly Observatory, HUN-REN Research Centre for Astronomy and Earth Sciences, MTA Centre of Excellence\\ 
H-1121 Budapest, Konkoly Thege Mikl\'os \'ut 15-17, Hungary\\
\and
E\"otv\"os Lor\'and University, Institute of Physics and Astronomy, H-1117, Budapest, Pázmány Péter sétány 1/a, Hungary
}

\abstract
{During the primary \textit{Kepler} mission, between 2009 and 2013, about 150,000 pre-selected targets were observed with a 29.42 minute-long cadence. However, a survey of background stars that fall within the field of view (FOV) of the downloaded apertures of the primary targets has revealed a number of interesting objects. In previous papers we have presented surveys of short period Eclipsing Binaries  and RR Lyrae stars}
{The current survey of the \textit{Kepler} background is concentrated on identifying longer-period eclipsing binaries and pulsating stars. These will be the subject of later papers. In the course of this survey, in addition to eclipsing binaries and pulsating stars, seven exoplanet candidates have been uncovered and in this paper we report on these candidates.}
{We use Lomb-Scargle (LS), light curve transit search and Phase Dispersion Minimisation (PDM) methods to reveal pixels that show significant periodicities, resulting in the identification of the seven exoplanet candidates. We have prepared the light curves for analysis using Pytransit software and cross matched the pixel coordinates with Gaia and other catalogues to identify the sources.}
{We identify seven Hot Jupiter exoplanet candidates with planet radii ranging from 0.8878 to 1.5174 $R_{Jup}$ and periods ranging from 2.5089 to 4.7918 days.}
{}


\maketitle

\section{Introduction} \label{sec:intro}

Since the first discovery of an exoplanet around 51 Peg, \citep{1995IAUC.6251....1M} and of the first transiting exoplanet in 1999, \citep{2000ApJ...529L..45C, 2000ApJ...529L..41H}, over 5600 exoplanets have been discovered. In the following decades, surveys such \textit{OGLE} \citep{OGLE1997}, \textit{CoRoT} \citep{CoRoT2009}, \textit{Kepler} \citep{borucki2010} and the ongoing \textit{TESS} \citep{TESS2015} have contributed to this dramatic rise in the number of known exoplanets.

The study of exoplanets is important for understanding how planetary systems form and evolve and for understanding whether and where life might exist elsewhere in the Universe. They additionally give clues to how our own Solar System was formed.

The \textit{Kepler} photometric space telescope was launched in 2009 into an Earth-trailing heliocentric orbit and was designed to detect exoplanets (and earth-analogues in particular) by observing around 150,000 target stars in a fixed 105 square degree area of the sky in the Cygnus, Lyra and Draco constellations. A $4\arcsec$ pixel size was used to provide a large full-well capacity and enable the high signal-to-noise ratio needed to detect earth-sized planets. The observations were made with a 29.42 minute (long) cadence over a period of 17 Quarters from 2009 to 2013 at which time a second reaction control wheel failed and this part of the mission was terminated. For telemetry bandwidth reasons, only pixels of the target stars and their immediate surroundings were downloaded, and analysis of these images has focused on the pixels within the optimal apertures of the target stars. However, a pixel-by-pixel analysis of these images reveals a variety of interesting objects in the background.

In previous papers, we have described findings of eclipsing binary stars (\cite{2021ApJS..256...11B} and \cite{2020CoSka..50..405F}) and RR Lyrae stars \citep{2022ApJS..260...20F}. The current survey was carried out with the primary objective of identifying longer-period eclipsing binary stars and pulsating stars, with periods up to approximately 90 days, and these will be the subjects of future papers. However, in the course of this survey, seven exoplanet candidates have been identified and these are the subject of this paper.

In Sect.~\ref{sec:processing} we discuss data processing, including light curve processing and search methods, candidate selection and identification of host stars. In Sect. \ref{sec:cand_analysis} we discuss the analysis of the candidates, including transit modeling, vetting, transit timing variation, stellar and planetary parameters, comparisons with known exoplanets and selection biases. In Sect.~\ref{sec:results1} we briefly discuss our findings and in Sect.~\ref{sec:summary} we give a brief summary of our work.

Note that the candidates are referred to in this paper by the \textit{Gaia} EDR3 catalogue reference of the host star, followed by the Kepler Input Catalog (KIC) reference of the aperture in which they were found in brackets, e.g. "Gaia DR3 2128959007378441472 (KIC 4459924)"

\section{Processing of \textit{Kepler} data} \label{sec:processing}
\subsection{Detrending procedure} \label{sub:detrend_proc}

The \textit{Kepler} observations were made quasi-continuously over a 4 year period. However, the data is provided in discrete Quarterly sets due to the 90-degree rolls required to maintain the correct orientation of the solar panels.
Our initial search was restricted to the Quarter 4 (Q4) observations, since this was the first relatively quiet Quarter. All the individual pixels belonging to long cadence Q4 KIC target apertures were used in the search.

The light curves were extracted for each individual pixel for each main target (denoted by KIC numbers) and detrended. This is a necessary precursor to generating the periodograms, as the drift in the light curves generates spurious frequencies in the area of interest.

The quarterly light curves are subject to discontinuities and so each light curve was split into segments which could be detrended separately. Each such segment was then detrended and normalised by fitting, and dividing by, a quadratic (one quadratic per segment) and the individual segments were then re-combined into a single light curve. A quadratic fit was used as some of the segments are relatively short and use of a higher order polynomial would likely remove any low frequency cyclic variation in such segments.

\subsection{Search procedure} \label{sub:search_proc}

After detrending, a low-resolution Lomb-Scargle (LS; \citealt{Lomb76} and \citealt{Scargle82}) algorithm with a frequency interval $\approx 0.01$ cycles/day was employed to generate a periodogram for each pixel light curve.

 Searching individual pixels allows targets not located on a KIC aperture to be identified as long as their PSFs overlap the aperture.

The resulting periodograms were searched for individual pixels showing a significant  dominant frequency less than 4 cycles/day (equivalent to a period of 6 hours for an exoplanet). 'Significant' means the amplitude is greater than 4 times the local background, that is: 
\begin{equation}
a_{d} - a_{ave} \geq 4 \times d_{rms}
\end{equation}
where $a_{d}$ is the dominant frequency amplitude, $a_{ave}$ is the average amplitude of the frequencies within $\pm 0.5$ c/d of the dominant frequency, and $d_{rms}$ is the RMS deviation from $a_{ave}$ of the frequencies within $\pm 0.5$ c/d of the dominant frequency. In many cases the dominant frequency was less than 0.5 cycles/day, and in these cases the background was measured between 0 and 1 cycle/day.

Even with detrending, the level of noise in the periodogram below approximately 0.13 cycles per day means that the effective upper period limit to the periodogram search is approximately 8 days for an exoplanet (16 days for an eclipsing binary). Hence, a second routine was used to search the light curves looking for significant dips in the flux. As an experiment, a simple search routine was developed rather than using the more usual BoxLeastSquares (BLS) approach. Testing showed this to be as sensitive as BLS analysis.

In this case, 'a significant dip' means an observation where the difference between the flux level and the average flux level is more than 3.5 times the RMS deviation of the observations, that is:

\begin{equation} \label{equ:lc_search}
f_{ave}- f_{d} \geq 3.5 \times f_{rms}
\end{equation}

where $f_{d}$ is the flux of a single observation, $f_{ave}$ is the average flux across the detrended light curve, and $f_{rms}$ is the RMS deviation from $f_{ave}$ of the flux across the light curve. A minimum of 3 such dips were required to select the light curve as a candidate. No requirement was made for the dips to be equally spaced, but subsequently, lightcurves and periodograms for all the candidates, found by either method, were visually inspected and false positives discarded.

Note that all of the exoplanet candidates in this paper were found with this second search method.

After checking for known objects, false positives (arising from bleeding, ghosting or crosstalk), rotational variables and duplicates (where a candidate PSF is detected in two KIC apertures), the individual pixel light curves for each remaining candidate were merged and detrended, using the procedure described above. 

For each candidate, a search was carried out on the relevant Q2 - Q16 pixel light curves to identify the pixels that fold at the same frequency as the identified candidates. Between 1 and 4 pixels were found for each candidate for each quarter (though in some quarters no pixels were found or there was no data for the KIC aperture in question). Quarters 0, 1 and 17 were discarded as they are truncated and give rise to a disproportionate number of frequencies that differ significantly from those obtained from Quarters 2 - 16. The resulting pixel light curves were then merged, detrended and normalised to provide a set of Quarterly light curves. These were then merged into a single Q2 - Q16 light curve.

\subsection{Exoplanet candidate selection} \label{sub:cand_sel_proc}

Initial candidate selection was based on the Q4 individual pixel light curves only. Light curves were classified as possible exoplanets if they met a number of criteria:
\begin{enumerate}
  \item The light curve should have the correct morphology, i.e. the continuum flux of the phase-folded light curve should be constant between transits. 
  \item The transit depth should be less than 2.5\%.
  \item There should be no apparent secondary eclipse.
  \item The host star should be unambiguously identifiable (The source identification process is described in Sect. \ref{sub:source_id}).
  \item The Gaia catalogue entry for the host star should provide sufficient data to allow the radius of the star (and thus the exoplanet candidate radius and orbital radius) to be estimated.
\end{enumerate}

Finally, after the quarterly light curves were combined into a single Q2 - Q16 light curve, only those candidates were retained that had the typical planetary U-shaped transits. 

\begin{figure}[!ht]
\centering
\includegraphics[width=\columnwidth]{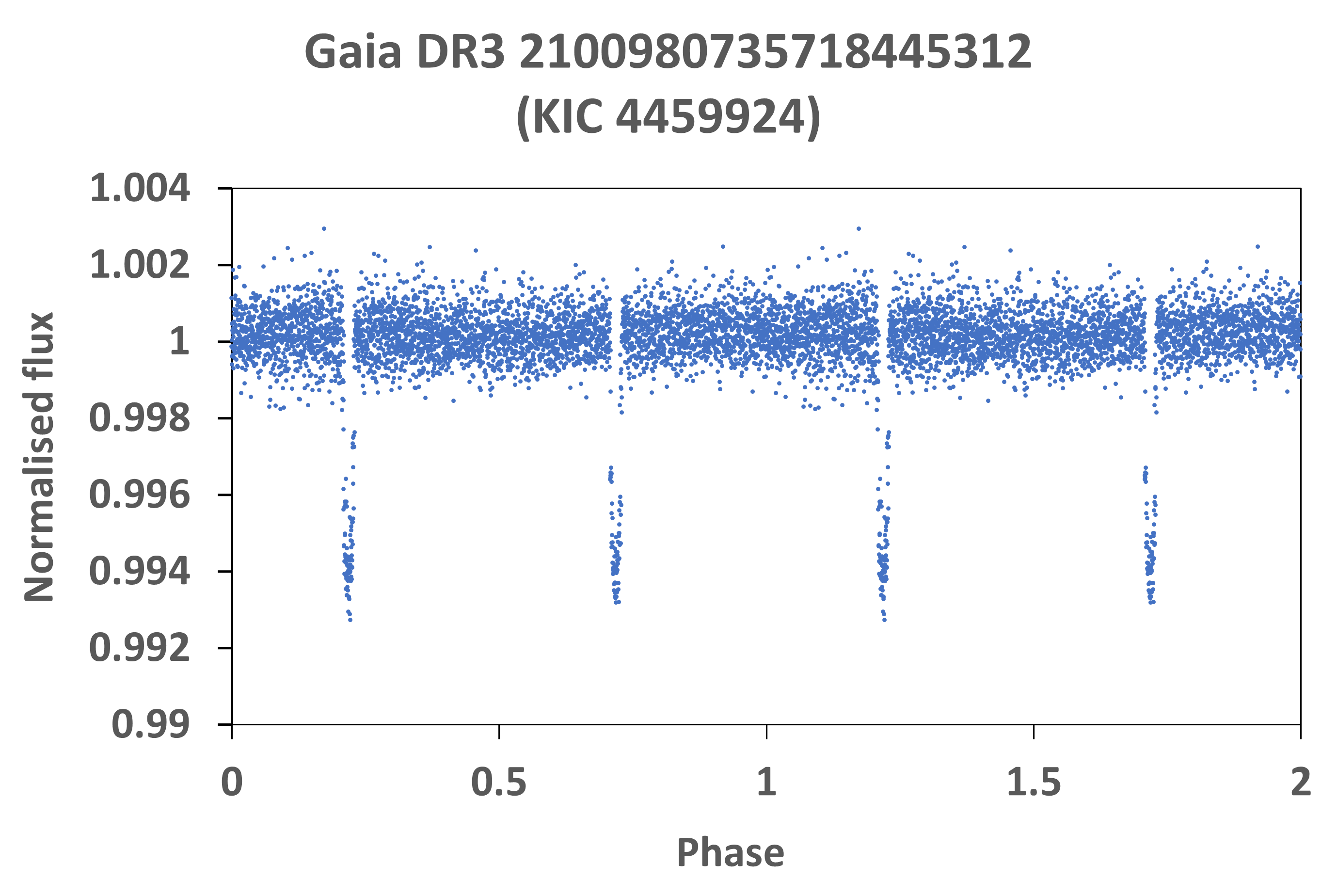}
\includegraphics[width=\columnwidth]{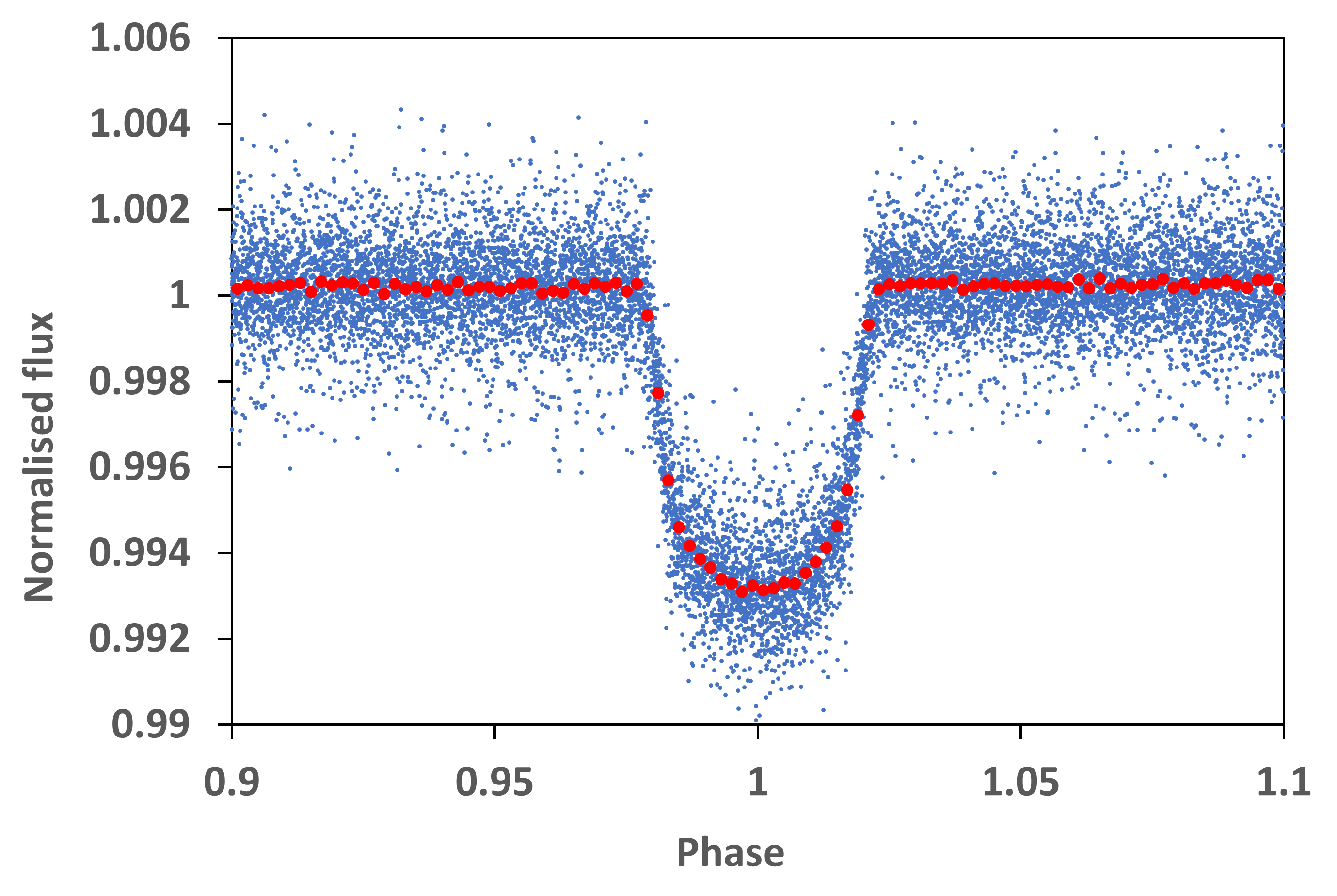}
\caption{(\textit{Top}) the Q4 light curve of the candidate found in the background of the KIC 4459924 aperture (before calculation of the zero epoch, $BJD_0$), phase folded as though it is an eclipsing binary. This meets the first three qualifying criteria for an exoplanet candidate. (\textit{Bottom}) the phase folded Q2 - Q16 light curve of the candidate showing the U shape typical of a planetary transit. Blue dots are individual observations and red circles are the phase bin mean fluxes.}
\label{fig:4459924_lc}
\end{figure}

An example of a qualifying folded light curve is given in Fig.~\ref{fig:4459924_lc}, showing the Q4 light curve found in the background of the KIC 4459924 aperture and the folded Q2 - Q16 light curve showing the typical U-shaped exoplanet transit.

Vetting of the selected candidates is described in Sect. \ref{sub:vetting}.

\subsection{\textbf{Light curve deblending}} \label{sub:light_deblend}

The candidate host stars all lie between approximately 5$\arcsec$ and 23$\arcsec$ from the (generally brighter) \textit{Kepler} main targets, and are thus subject to light contamination from these same main targets. The effect of such contamination is to reduce the apparent fractional transit depths and thus the apparent radii of the candidate exoplanets.

PSFmachine is a software package specifically designed to deblend Kepler light curves \citep{2023ascl.soft06056H} and we have employed this to generate deblended light curves for each candidate and each quarter. Unfortunately, a number of issues have arisen with this, including a failure to produce some light curves, a lack of reproducibility in the generated light curves and a lack of consistency across quarters, with quarterly light curves having quite wide variations in flux and transit depth. We speculate that this is due to the candidates mostly being located adjacent to, rather than on, their respective KIC apertures and the generally weak flux levels of the light curves.

In one extreme case, Gaia DR3 2085335681690420608 (KIC 9674230), PSFmachine was unable to generate any light curves at all, possibly because of a weak light curve: only one pixel in each quarter exhibits a usable signal.

However, where PSFmachine was able to generate light curves, it enabled verification of the source stars. One candidate was eliminated as PSFmachine was able to show that the incorrect source star had been selected.

Instead of using PSFmachine, the candidate light curves were obtained by selecting and merging the pixel light curves exhibiting the best signals, i.e. those with the deepest transit depths, and discarding those with lower transit depths, thus effectively minimising the light contamination from the main target.

\begin{figure}
\centering
\includegraphics[width=\columnwidth]{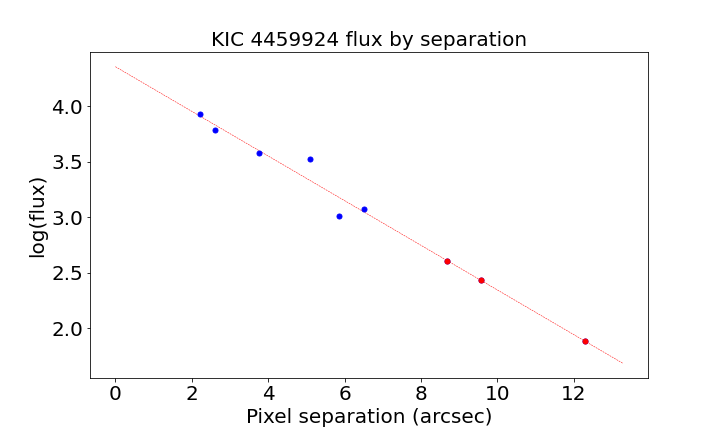}
\caption{This shows the Q14 main target average flux per pixel as a function of the separation (distance) of the pixel centre location from the main target location in arcsec (blue circles). The red line is the fit to log(flux), and the red circles indicate the extrapolation of the fitted line to the locations of the exoplanet candidate pixels.} 
\label{fig:4459924_cont}
\end{figure}

The residual main target contamination was then estimated for each quarterly light curve by plotting the distribution of the main target flux as a function of distance from the main target sky location. To a reasonable approximation, the main target flux decreases logarithmically with distance from the main target location.  Fig.~\ref{fig:4459924_cont} shows a typical example of this for the KIC 4459924 Q14 aperture. The estimated contamination obtained by this method is subtracted from the quarterly light curves for each candidate.

In some instances, particularly if the main target comprises only 2 or 3 pixels, this approach produces unrealistic results. In these cases, the contamination is taken from the corresponding quarters (i.e. quarter numbers that differ by 4, 8 or 12 from the quarter in question) or, if this is not possible, the quarterly light curve is discarded.

The contamination flux is quite significant, ranging from 10\% to 40\% of the total candidate flux.

Two of the candidate light curves, Gaia DR3 2100325529865507712 (KIC 4543171) and Gaia DR3 2079369937757906688 (KIC 9040849) have also been published by \cite{2023AJ....166..265M}, and we have used these as a means of verifying the light curves we have obtained. The results are shown in Sect. \ref{sub:trans_model}.

\subsection{Determination of orbital periods} \label{sub:orb_period}

\begin{figure}
    \centering
    \includegraphics[width=\columnwidth]{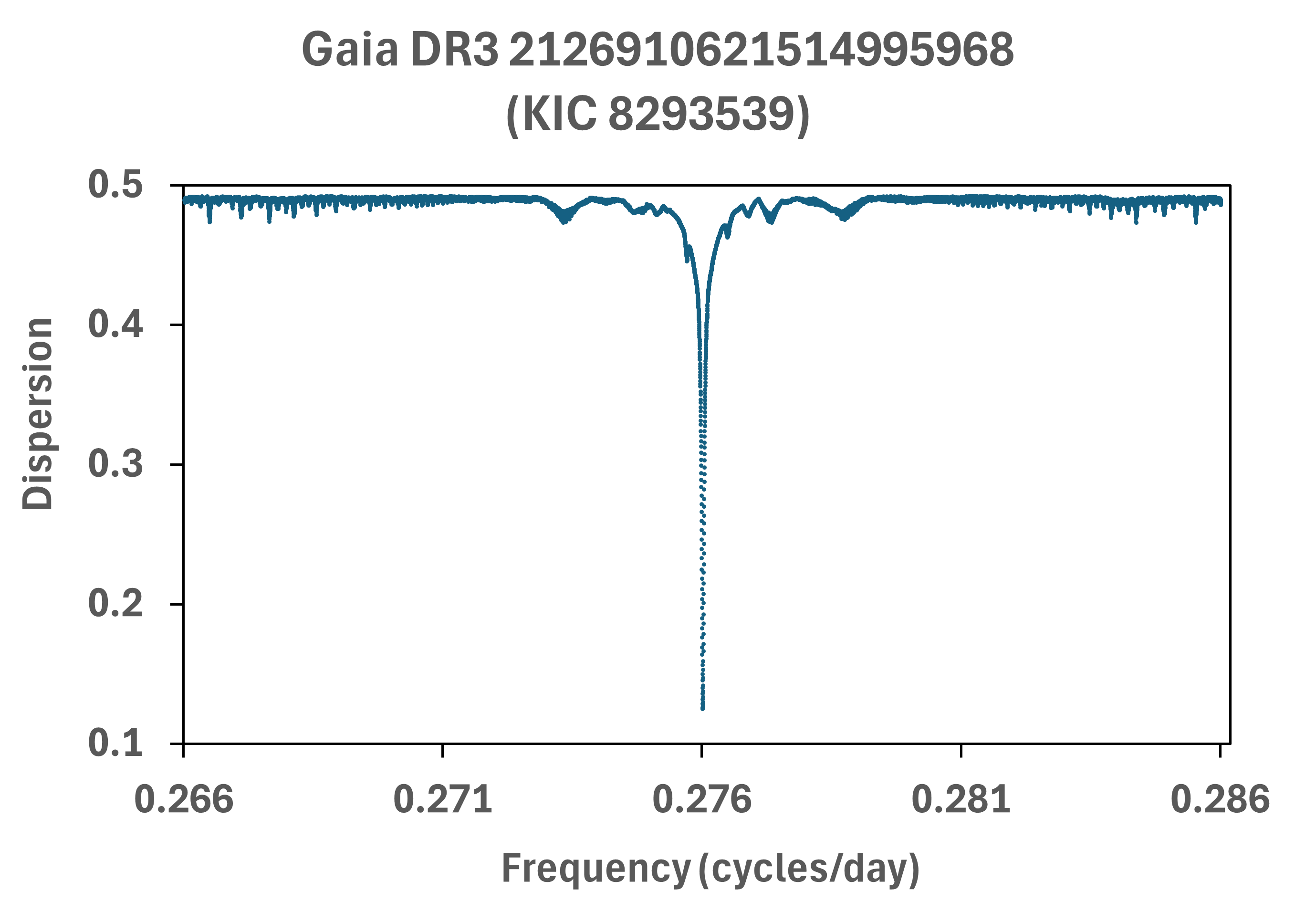}
    \includegraphics[width=\columnwidth]{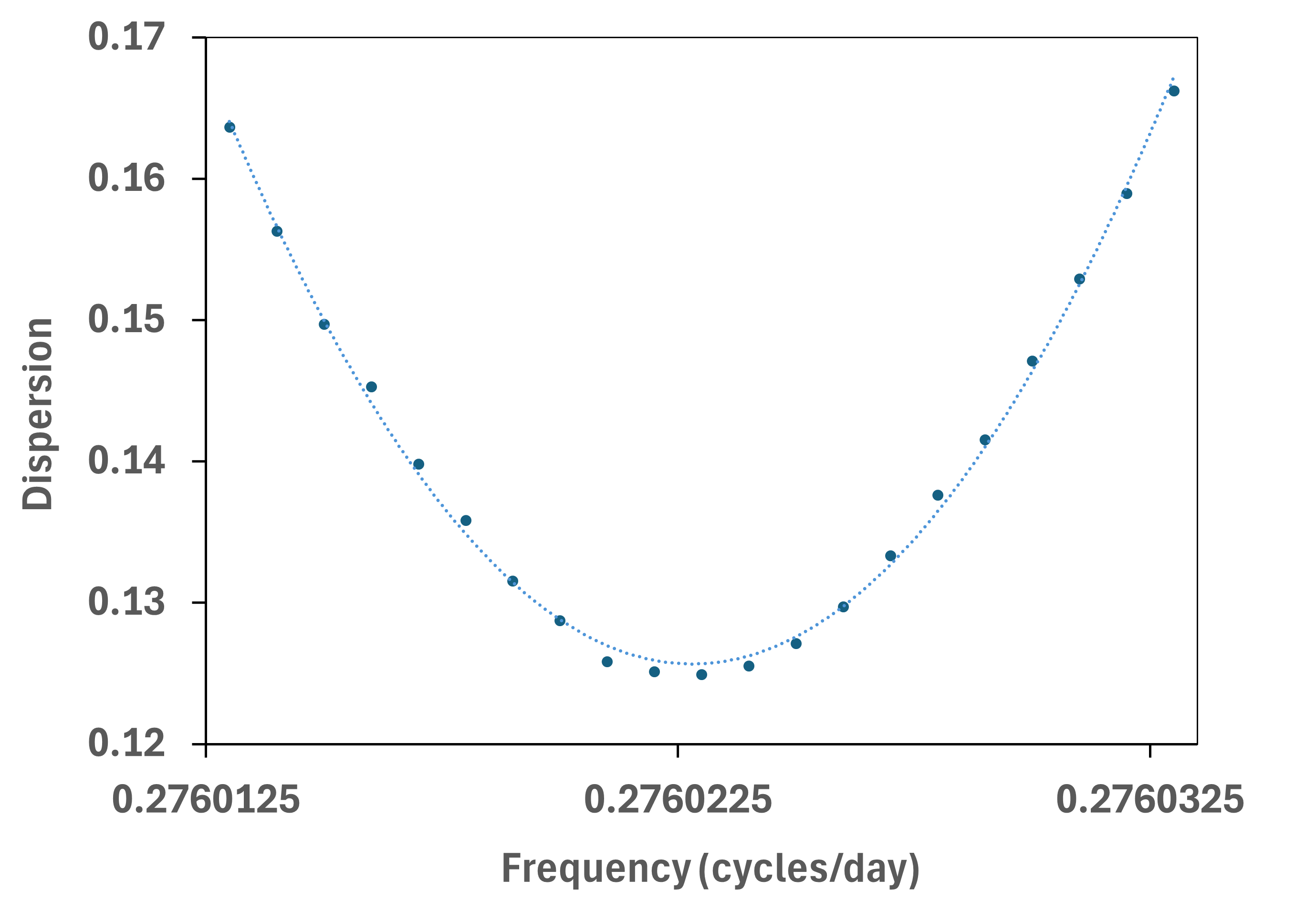}
    \caption{(\textit{Top}) this shows the variation of PDM dispersion with frequency for the candidate in the background of the KIC 8293539 aperture. (\textit{Bottom}) this shows the dispersion points around the minimum, fitted with a parabola which gives a frequency = 0.2760227 cycles/day (period = 3.6228904 days.)} 
    \label{fig:dispersion}
\end{figure}

An initial estimate of the period was made by the light curve search routine, using the intervals between successive transits. This was not always accurate due to spurious dips in the flux. The smallest period value was selected to allow for missing transits. The estimates obtained were progressively improved, using the method we previously employed for eclipsing binary stars \citep{2021ApJS..256...11B} as follows:
\begin{enumerate}
    \item For each Quarterly light curve, a Phase Dispersion Minimization (PDM) algorithm was employed to search for the frequency (in a range of \(\pm0.01\) cycles/day around the initially obtained frequency) which yields the minimum dispersion (scatter) for the phase-folded light curve, that is, the minimum value of:
    
\begin{equation}
\sum_{i=1}^{n}\sum_{j=1}^m(O_{ij}-E_i)^{2}/E_i
\end{equation}

where $n$=30 is the number of phase bins, $O_{ij}$ is the ${j}^{th}$ observed flux value for the ${i}^{th}$ bin and $E_i$ is the average flux value for the ${i}^{th}$ bin. The average frequency thus obtained was used as the starting point for the following step. 

   \item The same PDM algorithm was employed, but with 500 phase bins and using the Q2 - Q16 combined light curve to search for the best frequency in a range of \(\pm0.01\) cycles/day around the starting frequency. A fine adjustment was then made to the resulting frequency by fitting a parabola to the dispersion curve close to the minimum value. Fig.~\ref{fig:dispersion} shows an example of a dispersion variation by frequency. The \textit{top} chart shows a typical dispersion curve, with a clearly defined minimum. The fit to the fine dispersion curve is shown in the \textit{bottom} chart. 
   \item A further small adjustment was made to the period obtained using the Transit Timing Variation (TTV) measurements described in Sect. \ref{sub:ttv}.
\end{enumerate}

\subsection{Determination of zero epoch} \label{sub:zero_epoch}
\begin{figure}
    \centering
    \includegraphics[width=\columnwidth]{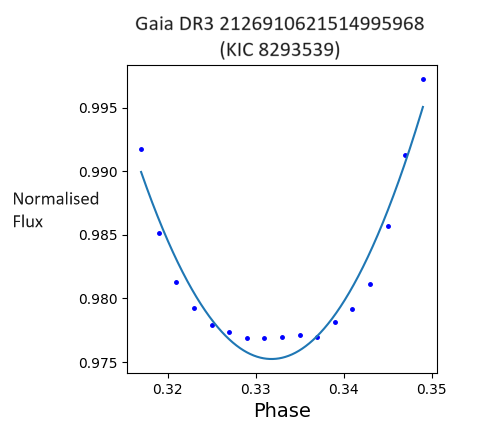}
    \caption{This figure shows the parabolic fit to a phase folded and binned transit light curve used to calculate the initial value of $BJD_0$ for the candidate.}
    \label{fig:t0_fit}
\end{figure}

The phase-folded and binned light curves were used to determine the zero epoch of the exoplanet candidates. A quadratic was fitted to the transit and the lowest point of this was assigned a phase value of zero. The last date before the start of the light curve corresponding with this phase was assigned to the zero epoch, $BJD_0$. However, it may be seen from Fig. \ref{fig:t0_fit} that the transit shape does not lend itself very well to a parabolic fit and this initial value of $BJD_0$ must be corrected. This was done using the transit modelling described in Sect. \ref{sub:trans_model} with a further correction obtained from the Transit Timing Variation (TTV) analysis described in Sect. \ref{sub:ttv}

These values are shown in \href{https://zenodo.org/records/14618891?token=eyJhbGciOiJIUzUxMiJ9.eyJpZCI6ImE0YmE1ZTMyLTUyNDUtNDZhYi05NmI1LTUzYWM4NTgzYTVkYiIsImRhdGEiOnt9LCJyYW5kb20iOiIzNWFmZDRhMjExNjNlM2RjMjI4YzMyOGIyMTVmZjFjMSJ9.Q2GpAqZLxefIloSYfaJyUSOC8IenSx6kd6WeCiexA4oxUoyA6pfIg5TH16bcK2fUa_hCqroeC1LLBn8wSbaYzw}{Appendix A: Table 4}.

\subsection{Identification of host stars} \label{sub:source_id}

The potential host stars for the exoplanet candidates were found from the right ascension (RA) and declination (DEC) of the main pixel, that is, the brightest pixel from the matching set. Candidates for the host stars are those at, or close to, the main pixel coordinates. In the majority of cases, the pixels found lie on the edge of the KIC aperture mask and the centre of the candidate PSF lies outside the mask. As a result, in some cases, there were several possible host stars for a single exoplanet candidate. These were discarded since it is not possible to reliably estimate the candidate planetary parameters.

\begin{figure}
\centering
\includegraphics[width=\columnwidth]{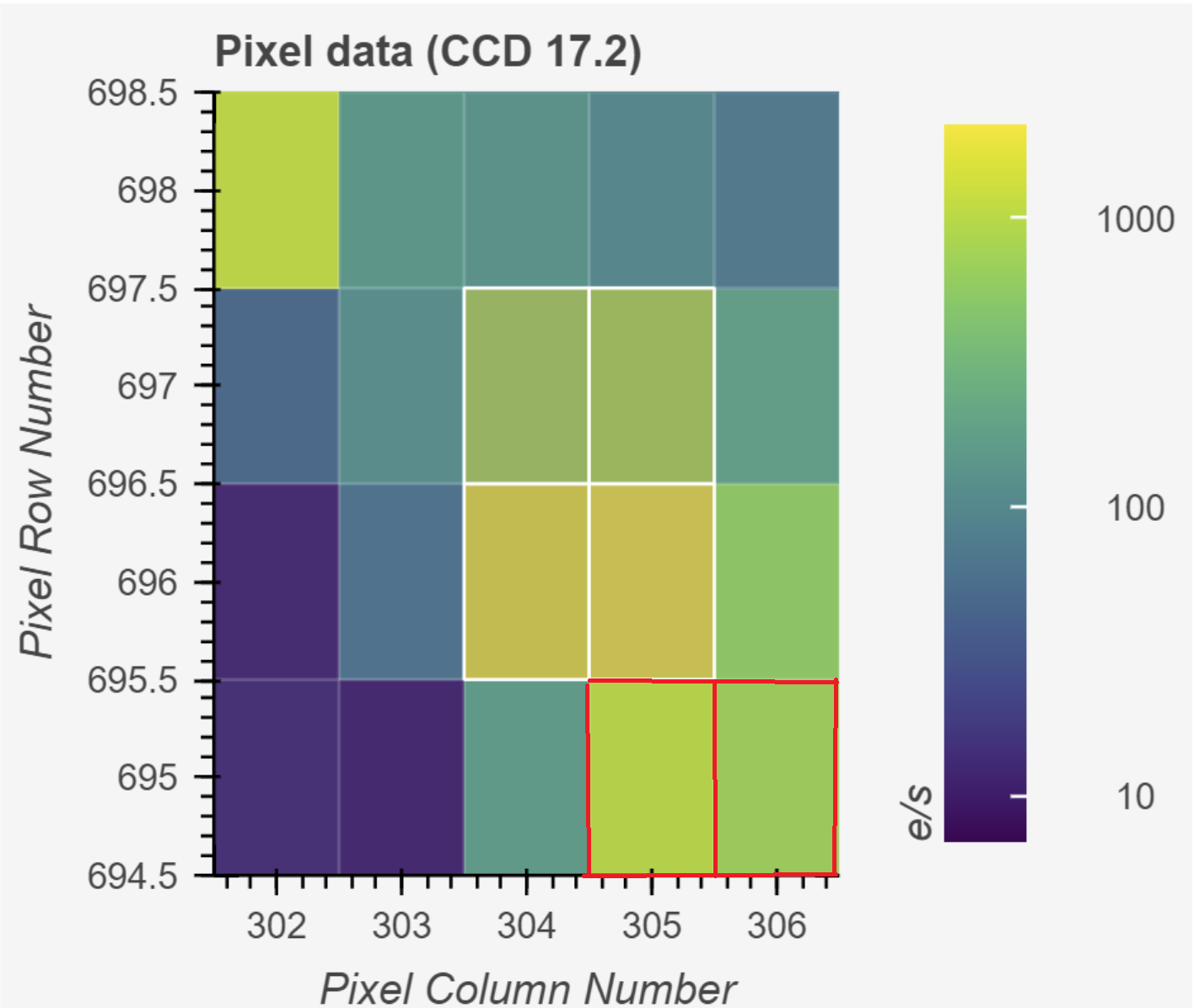}
\caption{KIC 8175131 aperture mask with the exoplanet candidate pixels outlined in red. The four bright pixels used in the \textit{Kepler} pipeline to extract the main target light curve are outlined in white. (source: Lightkurve)}
\label{fig:8175131_pix}
\end{figure}

\begin{figure}
  \centering
  \includegraphics[width=\columnwidth]{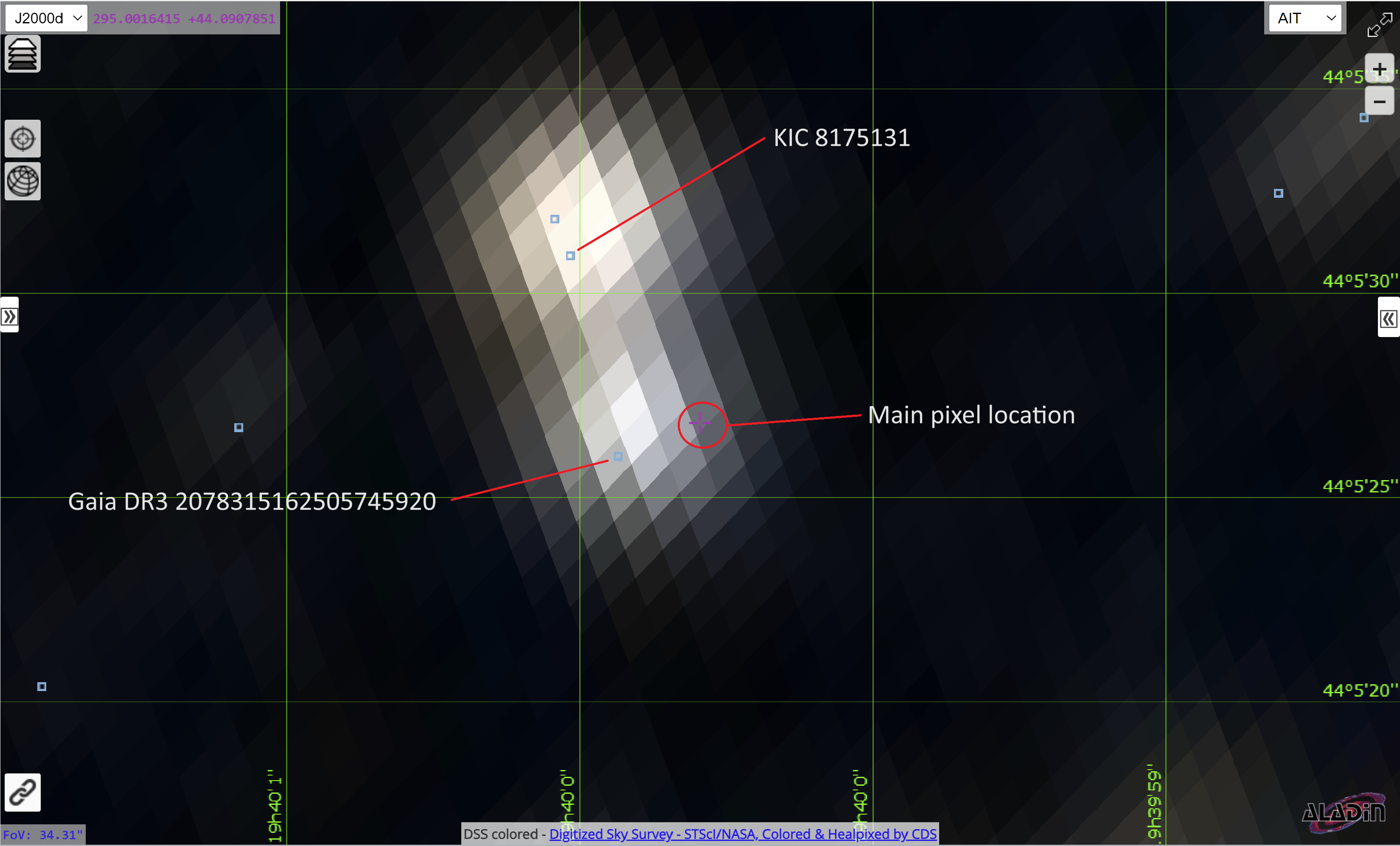}
  \caption{Aladin sky map showing the area around KIC 8175131. The main pixel location for the exoplanet candidate and the host star (Gaia DR3 2078315162505745920) are marked. The FOV is $34.31\arcsec$ across.}
  \label{fig:8175131_aladin}
\end{figure}

An example is shown in Figs.~\ref{fig:8175131_pix} and \ref{fig:8175131_aladin}. Fig.~\ref{fig:8175131_pix} shows the pixel-by-pixel aperture mask for KIC 8175131. The four bright pixels used in the \textit{Kepler} pipeline to extract the main target light curve are outlined in white and the two pixels outlined in red are where the exoplanet candidate light curve was detected. The RA and DEC for the main pixel (row 695, column 305) are obtained using Astropy and the \textit{Gaia} catalogue searched for all stars within 12\arcsec (3 pixels) of this location. These are then checked with Aladin to find the actual source. Fig.~\ref{fig:8175131_aladin}, obtained from Aladin, shows the area of sky around this location, with the main pixel location, KIC 8175131 and the host star (Gaia DR3 2078315162505745920) marked.

The identified \textit{Gaia} catalogue references for the candidates are listed in \href{https://zenodo.org/records/14618891?token=eyJhbGciOiJIUzUxMiJ9.eyJpZCI6ImE0YmE1ZTMyLTUyNDUtNDZhYi05NmI1LTUzYWM4NTgzYTVkYiIsImRhdGEiOnt9LCJyYW5kb20iOiIzNWFmZDRhMjExNjNlM2RjMjI4YzMyOGIyMTVmZjFjMSJ9.Q2GpAqZLxefIloSYfaJyUSOC8IenSx6kd6WeCiexA4oxUoyA6pfIg5TH16bcK2fUa_hCqroeC1LLBn8wSbaYzw}{Appendix A: Table 1}together with their R.A. and Declination and any alias identifiers. 

\section{Candidate analysis} \label{sec:cand_analysis}

\subsection{Transit modeling} \label{sub:trans_model}

Modelling of the exoplanet candidate light curves was performed using the PyTransit software package \citep{Parviainen_2015}.

The software incorporates log posterior functions (LPFs) for transit modelling and parameter estimation. These  store the observations, model priors etc. to create a basis for Bayesian parameter estimation from transit light curves and and also contain methods for posterior optimisation and Markov Chain Monte Carlo (MCMC) sampling.

PyTransit implements the quadratic limb darkening model of \cite{2002ApJ...580L.171M} and \cite{2020AJ....159..123A}, so that, in general terms:

\begin{equation}
    F = \frac{I_* - \overline{I}(k,g)A(k,g)}{I_*}
\end{equation}

where \textit{F} is the observed fractional flux, $I_*$ is the stellar flux, $\overline{I}(k,g)$ is the average limb darkening-corrected flux covered by the planet, and $A(k,g)$ is the planet-star intersection area. $\overline{I}$ and $A$ are expressed as functions of the planet-star radius ratio, $k$, and the grazing factor, $g$, which is equal to $b/(1+k)$ where $b$ is the impact parameter.

The principal parameters calculated by the model are (using the PyTransit notation):\\
$b$: the impact parameter, i.e. the projected height of the transit above the stellar midline as a fraction of the stellar radius.\\
$k$: the ratio of the planet radius to the stellar radius\\
$u$ and $v$: the quadratic model limb darkening coefficients\\
$a$: the ratio of the orbital radius to the stellar radius\\
$i$: the orbital inclination.\\
$t14$: the total duration of the transit.\\
$t23$: the duration of the complete transit.\\
$e$: the orbital eccentricity,\\
$w$: the argument of periastron.\\

The model was executed using the phase-folded individual observations of the normalized light curve, i.e. with the orbital period set to 1.0 days. Given the probabilistic nature of the modelling, the model was executed twice for each candidate, with the same light curve data, to ensure consistent results were obtained, i.e. the differences in the transit parameters between executions were much smaller than the standard deviations of the parameters. In the worst case, the ratio of the difference between the values obtained from the two runs to the standard deviation value was 0.1142. This was for the value of $k$, for Gaia DR3 2100980735718445312 (KIC 4459924).

The results are shown in  \href{https://zenodo.org/records/14618891?token=eyJhbGciOiJIUzUxMiJ9.eyJpZCI6ImE0YmE1ZTMyLTUyNDUtNDZhYi05NmI1LTUzYWM4NTgzYTVkYiIsImRhdGEiOnt9LCJyYW5kb20iOiIzNWFmZDRhMjExNjNlM2RjMjI4YzMyOGIyMTVmZjFjMSJ9.Q2GpAqZLxefIloSYfaJyUSOC8IenSx6kd6WeCiexA4oxUoyA6pfIg5TH16bcK2fUa_hCqroeC1LLBn8wSbaYzw}{Appendix A: Table 2}, these being the average values of the two runs for each candidate. Note that, in all cases, the orbital eccentricity was negligibly small (of the order of 10\textsuperscript{-10}) and is assumed to be zero. Thus $e$ and $w$ have been omitted from the table. 

It is noticeable that the sum of the impact parameter (\textit{b}) and the planet-star radius ratio (\textit{k}) is less than one in all cases, indicating that the transits are all full, and there are no grazing transits. 

In \href{https://zenodo.org/records/14618752?token=eyJhbGciOiJIUzUxMiJ9.eyJpZCI6ImY3OGNlYWNhLTBjZDQtNGIyYy1iNTJkLWFmYjk2NjJmZjRiNCIsImRhdGEiOnt9LCJyYW5kb20iOiIyMDVlNDcyNjJmMDQzMDMyOTllYzcwNDhjYWRiZTY5MSJ9.mhvK6uGYJ-4knrKz6fhhs-hvkdgU14SmEjqqxtKDG9uhognHf9vuCuarGdEaMAmj8f6USvVjDWkWU8Vqv1XplQ} {Appendix B}, we show the light curves obtained from the model. Each shows the individual observations and the model light curve overlaid with the phase-bin mean fluxes. The light curves all show the characteristic shape of a full, rather than grazing, transit, lending support to the calculated planetary radii.

In addition to the planetary and orbital parameters, the model identifies discrepancies between the zero phase point in the observations and the fitted model, and these discrepancies provide a correction to the value of $BJD_0$. These corrections, and subsequent corrections identified by TTV analysis, are shown in \href{https://zenodo.org/records/14618891?token=eyJhbGciOiJIUzUxMiJ9.eyJpZCI6ImE0YmE1ZTMyLTUyNDUtNDZhYi05NmI1LTUzYWM4NTgzYTVkYiIsImRhdGEiOnt9LCJyYW5kb20iOiIzNWFmZDRhMjExNjNlM2RjMjI4YzMyOGIyMTVmZjFjMSJ9.Q2GpAqZLxefIloSYfaJyUSOC8IenSx6kd6WeCiexA4oxUoyA6pfIg5TH16bcK2fUa_hCqroeC1LLBn8wSbaYzw}{Appendix A: Table 4}

Two of the candidate light curves, Gaia DR3 2100325529865507712 (KIC 4543171) and Gaia DR3 2079369937757906688 (KIC 9040849) have also been published by \cite{2023AJ....166..265M}, and we have modeled these with Pytransit as a means of assessing the reasonability of our light curves. The comparison of the principal Pytransit parameters ($i$, $k$ and $a$) is shown in \href{https://zenodo.org/records/14618891?token=eyJhbGciOiJIUzUxMiJ9.eyJpZCI6ImE0YmE1ZTMyLTUyNDUtNDZhYi05NmI1LTUzYWM4NTgzYTVkYiIsImRhdGEiOnt9LCJyYW5kb20iOiIzNWFmZDRhMjExNjNlM2RjMjI4YzMyOGIyMTVmZjFjMSJ9.Q2GpAqZLxefIloSYfaJyUSOC8IenSx6kd6WeCiexA4oxUoyA6pfIg5TH16bcK2fUa_hCqroeC1LLBn8wSbaYzw}{Appendix A: Table 3}. These are in reasonable agreement and lend confidence to our candidate light curves.

\subsection{Candidate vetting} \label{sub:vetting}

\subsubsection{Odd-even transit comparison} \label{subsub:odd_even_tran}

The odd-numbered and even-numbered transits for each candidate were subjected to separate PyTransit fits to ensure that the transit depths were the same in each case. A significant difference would indicate that the candidate is an eclipsing binary rather than an exoplanet. The results are shown in \href{https://zenodo.org/records/14618891?token=eyJhbGciOiJIUzUxMiJ9.eyJpZCI6ImE0YmE1ZTMyLTUyNDUtNDZhYi05NmI1LTUzYWM4NTgzYTVkYiIsImRhdGEiOnt9LCJyYW5kb20iOiIzNWFmZDRhMjExNjNlM2RjMjI4YzMyOGIyMTVmZjFjMSJ9.Q2GpAqZLxefIloSYfaJyUSOC8IenSx6kd6WeCiexA4oxUoyA6pfIg5TH16bcK2fUa_hCqroeC1LLBn8wSbaYzw}{Appendix A: Table 7}. For each candidate we show the difference between the odd- and even-numbered transit depths and the full light curve transit depth expressed as a fraction of the $1\sigma$ uncertainty in the full light curve transit depth. In all cases, the transit depth differences fall within the $1\sigma$ range.

The odd- and even-numbered Pytransit model fits are also shown graphically in \href{https://zenodo.org/records/14618766?token=eyJhbGciOiJIUzUxMiJ9.eyJpZCI6IjczMzliOGEyLWYyOGUtNDliZi1hYTM5LTk0Y2EwNjhjMmY1ZCIsImRhdGEiOnt9LCJyYW5kb20iOiI4YTg0YThlYTM1YzcyNDViYzU0ODdjNmVlMTQwNWI5ZiJ9.H63XLMFfgfqgfXsH2vTl-duISKPQShy3cOD4KE89q_os5GVzPFybxrTMgEGpARfGEKt5D5sJ-bo0GkGF1uGOAg}{Appendix C}. For each candidate, we show the model fit to the odd transits and even transits, the minimum flux obtained from the full light curve and the estimated $1\sigma$ error in the minimum flux.

\subsubsection{Centroid vetting}\label{subsub:centroid} 

Centroid vetting is the means by which a candidate star is checked to be the true source of an interesting signal, rather than being contaminated by some background star. A significant centroid offset is evidence that the selected candidate is a false positive.

For this purpose we have employed the Vetting python package \citep{Hedges_2021} which was constructed specifically for \textit{Kepler} light curves and carries out in-transit and out-of-transit tests to ensure that a light curve signal is centred on the selected candidate.

The test was run for each candidate and quarter, and generally confirm the candidate host stars. However, there are limitations that prevent Vetting from analysing some light curves:

\begin{enumerate}
\item{The pixel mask for the candidate cannot be in a 1xn format. Thus a minimum of 3 pixels, which must not be in a single row or column, is required for the test to work. For a number of candidates and quarters, only 1 or 2 pixels are available.}
\item{The KIC aperture must not be too crowded.}
\end{enumerate}

In cases where centroid testing could be carried out, i.e. there were sufficient pixels and the KIC aperture was not too crowded, the candidates were mostly confirmed to be the source. However, there were exceptions:

\begin{enumerate}
\item{Gaia DR3 2078315162505745920 (KIC 8175131): offsets were detected in Q8 and Q16, but not in Q4 and Q12. The same pixel mask was used in each case. There were insufficient pixels in the other quarters.}
\item{Gaia DR3 2126910621514995968 (KIC 8293539): offsets were detected in Q3 and Q7, but not in Q4, Q8, Q12 and Q16. There were insufficient pixels in other quarters.}
\item{Gaia DR3 2079369937757906688 (KIC 9040849): the KIC aperture is too crowded. No testing could be carried out.}
\item{Gaia DR3 2085335681690420608 (KIC 9674230): There was only one relevant pixel in each of the quarters. No testing could be carried out.}
\end{enumerate}

\begin{figure}
\centering
\includegraphics[width=\columnwidth]{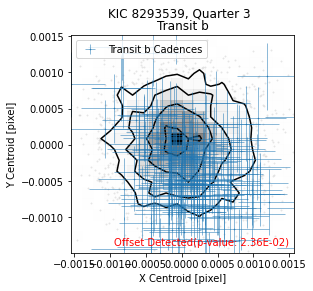}
\includegraphics[width=\columnwidth]{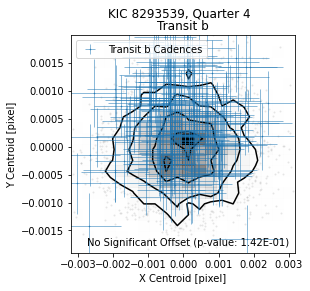}
\caption{This figure shows the Vetting output for Gaia DR3 2126910621514995968 (KIC 8293539) Q3 (\textit{top}) where an offset has been detected and Q4 (\textit{bottom}) where no offset has been detected.} 
\label{fig:centroid_vetting}
\end{figure}

The output from Vetting is illustrated in Fig. \ref{fig:centroid_vetting}, showing the output for Gaia DR3 2126910621514995968 (KIC 8293539) Q3, where an offset was detected and Q4 where no offset was detected.

\subsection{Transit timing variation (TTV)} \label{sub:ttv}

In a multi-planet system the gravitational effects of the planets on each other may perturb the timings of their transits, and by measuring these timing variations in one planet, it may be possible to infer the existence of other planets. See, for example, \cite{Ballard_2011}. Typically, a cyclic variation in the transit timing would indicate the presence of another planet.

We have carried out such a TTV analysis on the candidate light curves and, in fact, no such variations were observed; thus we have found no evidence for more planets in any of the systems examined.

\begin{figure}
\centering
\includegraphics[width=\columnwidth]{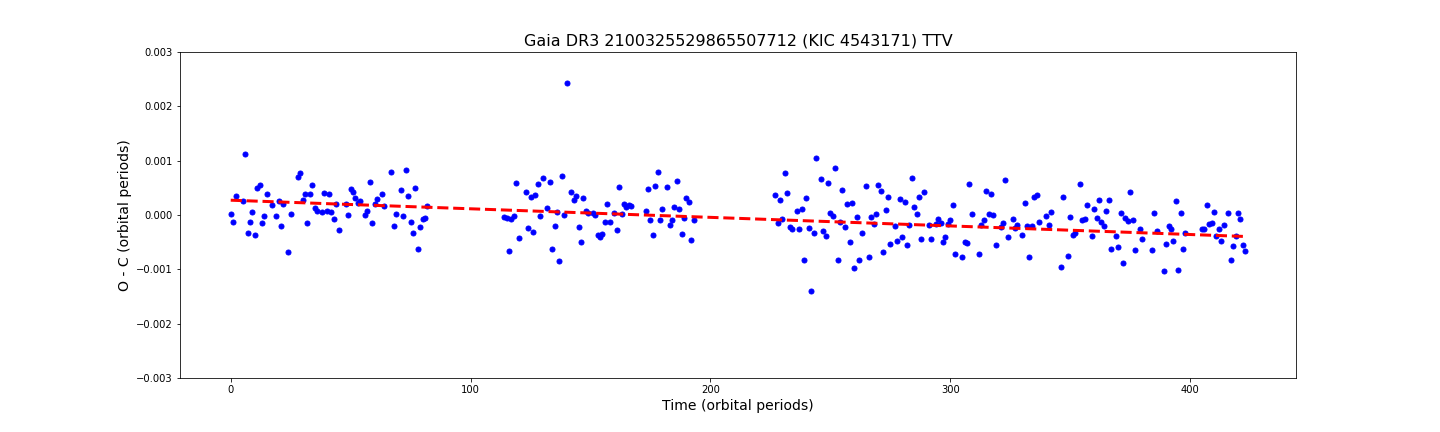}
\caption{This chart shows the initial TTV analysis for Gaia DR3 2100325529865507712 (KIC 4543171), showing the Observed - Calculated (O - C) values of the transit times before the $BJD_0$ and period corrections were applied. The gradient of the red trend line = -1.577E-06 and the average variation = -7.585E-05. Note that the time values are in orbital periods (= 3.2630113 days).}
\label{fig:ttv_corr}
\end{figure}

However, all of the TTV analyses showed a small linear trend and small non-zero values for the average variation. These are assumed to arise from errors in the orbital period and $BJD_0$. Corrections have been made to the period using the linear trend gradient and to $BJD_0$ using the average timing variation. After re-running the TTV with the corrected values, these discrepancies are mostly removed. Fig. \ref{fig:ttv_corr} shows a TTV chart before corrections were applied to $BJD_0$ and the orbital period, clearly showing the trend. The trend line has a gradient of -1.577E-06 and the average variation = -7.585E-05 (equivalent to a $BJD_0$ correction of -2.475E-04 days). 

The results of the corrections are shown in \href{https://zenodo.org/records/14618891?token=eyJhbGciOiJIUzUxMiJ9.eyJpZCI6ImE0YmE1ZTMyLTUyNDUtNDZhYi05NmI1LTUzYWM4NTgzYTVkYiIsImRhdGEiOnt9LCJyYW5kb20iOiIzNWFmZDRhMjExNjNlM2RjMjI4YzMyOGIyMTVmZjFjMSJ9.Q2GpAqZLxefIloSYfaJyUSOC8IenSx6kd6WeCiexA4oxUoyA6pfIg5TH16bcK2fUa_hCqroeC1LLBn8wSbaYzw}{Appendix A: Table 4}.

\subsection{Host star parameters} \label{sub:host_star_par}

The \textit{Gaia} catalogue provides values of apparent magnitude, $m_{G*}$, effective temperature, $T_*$ and gravity, $log(g_*)$ of the candidate host stars. In addition, we have taken the photogeometric distance, $d$, to each star as calculated by \cite{2021AJ....161..147B}. Together these allow estimations to be made of the stellar absolute magnitude, $M_{G*}$, luminosity, $L_*$, radius, $R_*$, and mass, $M_*$, using the standard equations:

\begin{equation}
M_{G*} = m_{G*}-5log_{10}\left(d/10\right)
\end{equation}

\begin{equation}
\frac {L_*} {L_\odot} = 10^{0.4\left(M_{G\odot} - M_{G*}\right)} 
\end{equation}

\begin{equation}
\frac {R_*} {R_\odot} = \left(\frac {T_\odot} {T_*}\right)^2.\left(\frac {L_*} {L_\odot}\right)^{0.5}  
\end{equation}

\begin{equation}
\frac {M_*} {M_\odot} = \left(\frac {R_*} {R_\odot}\right)^2\cdot10^{\left[log\left(g_*\right) - log\left(g_\odot\right)\right]}
\end{equation}

With these, estimates of the absolute (rather than relative) values of the exoplanet candidate planetary and orbital radii have been obtained. In addition, from the stellar mass together with the orbital period, $P_P$, the orbital radius, $a_P$, in A.U. may be estimated independently of the PyTransit model, using the standard equation:

\begin{equation}
a_P = \left[\left(\frac{P_P}{P_\oplus}\right)^2 \cdot \left(\frac{M_*}{M_\odot}\right)\right]^{1/3}
\end{equation}

The stellar parameters derived are shown in Table \href{https://zenodo.org/records/14618891?token=eyJhbGciOiJIUzUxMiJ9.eyJpZCI6ImE0YmE1ZTMyLTUyNDUtNDZhYi05NmI1LTUzYWM4NTgzYTVkYiIsImRhdGEiOnt9LCJyYW5kb20iOiIzNWFmZDRhMjExNjNlM2RjMjI4YzMyOGIyMTVmZjFjMSJ9.Q2GpAqZLxefIloSYfaJyUSOC8IenSx6kd6WeCiexA4oxUoyA6pfIg5TH16bcK2fUa_hCqroeC1LLBn8wSbaYzw}{Appendix A: Table 5}. The planetary radius estimates and both orbital radius estimates are included in \href{https://zenodo.org/records/14618891?token=eyJhbGciOiJIUzUxMiJ9.eyJpZCI6ImE0YmE1ZTMyLTUyNDUtNDZhYi05NmI1LTUzYWM4NTgzYTVkYiIsImRhdGEiOnt9LCJyYW5kb20iOiIzNWFmZDRhMjExNjNlM2RjMjI4YzMyOGIyMTVmZjFjMSJ9.Q2GpAqZLxefIloSYfaJyUSOC8IenSx6kd6WeCiexA4oxUoyA6pfIg5TH16bcK2fUa_hCqroeC1LLBn8wSbaYzw}{Appendix A: Table 6}

\subsection{Comparison with confirmed planets} \label{sec:NEA_compare}

\begin{figure}
\centering
\includegraphics[width=\columnwidth]{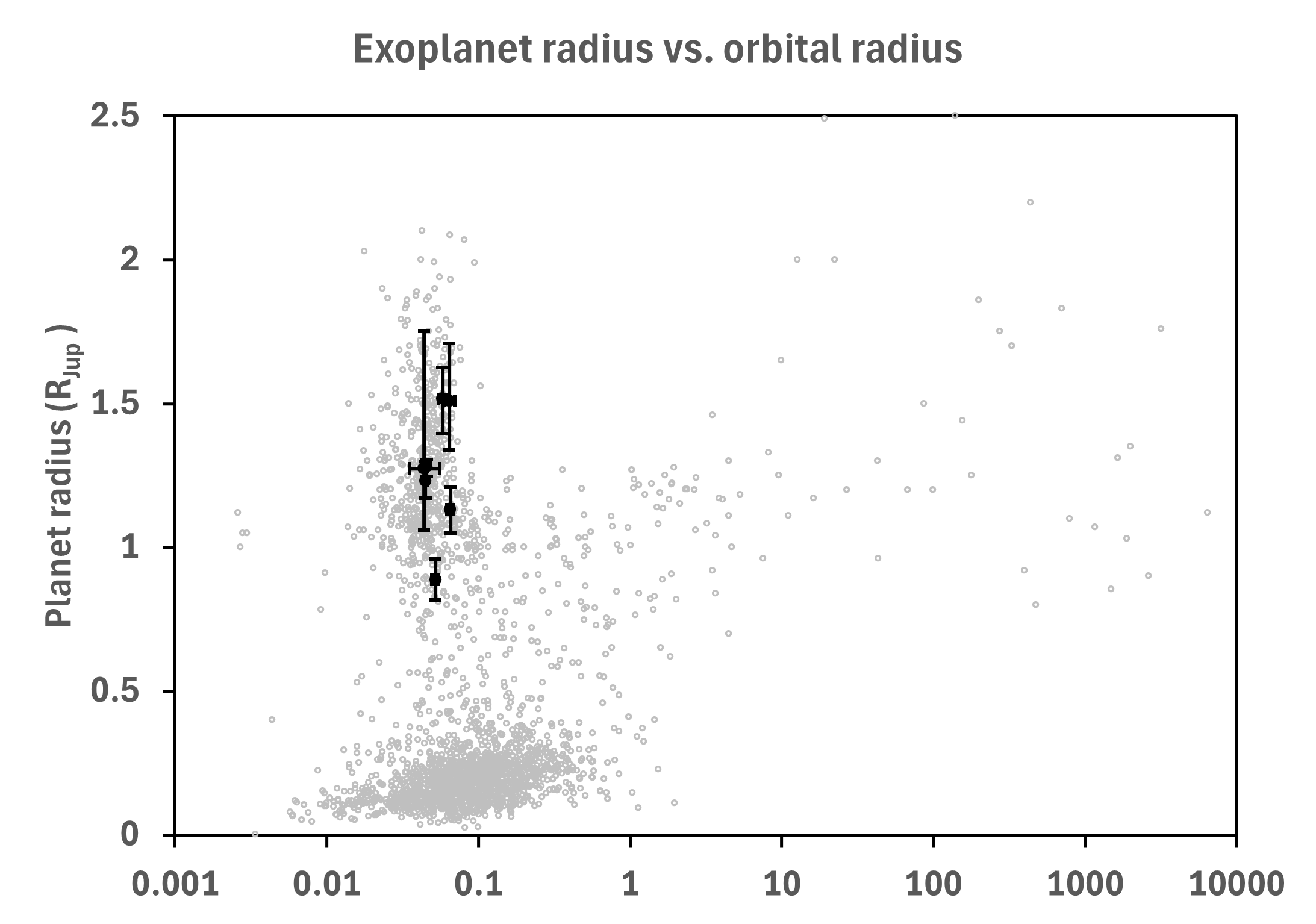}
\includegraphics[width=\columnwidth]{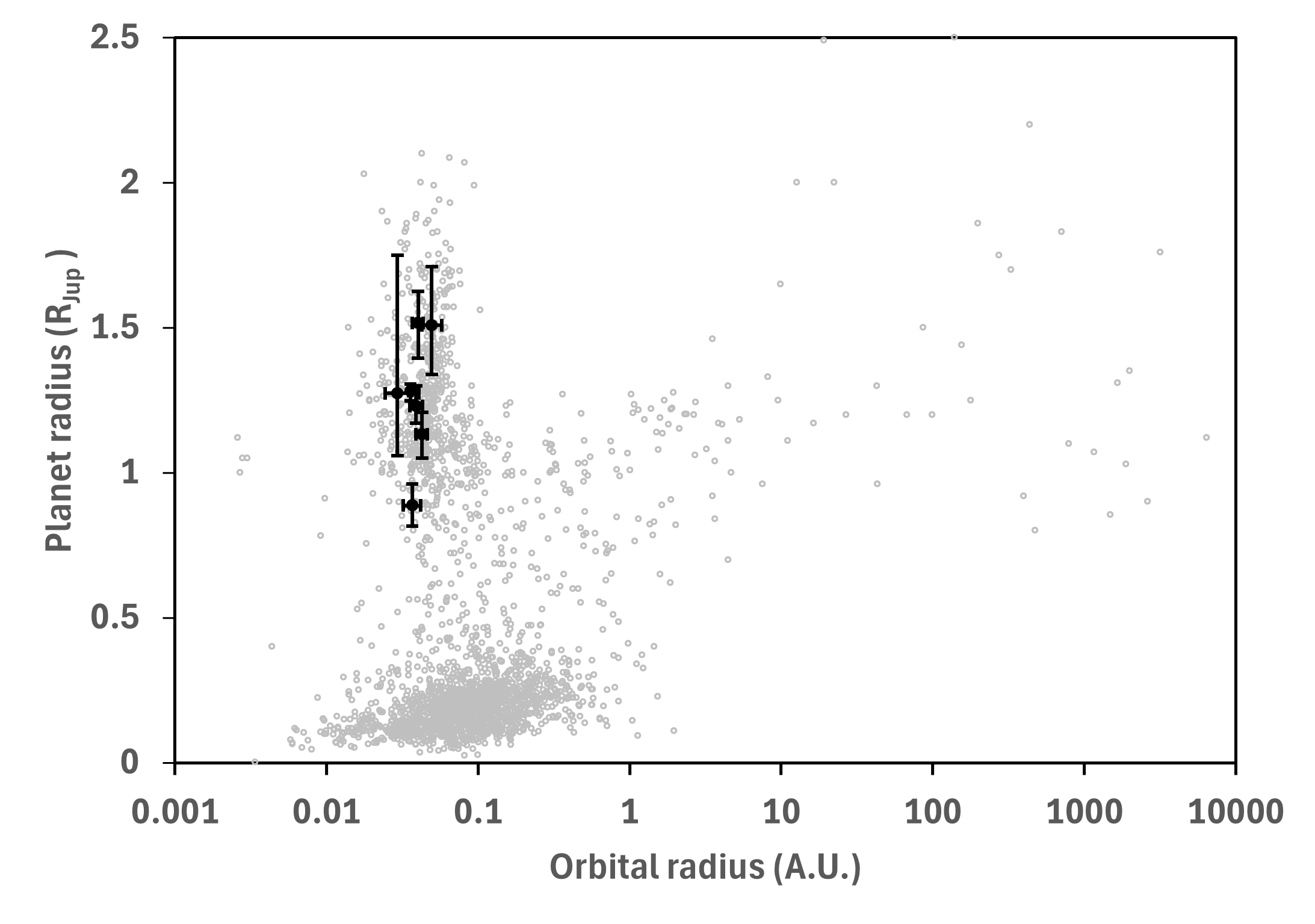}
\caption{This figure shows the confirmed exoplanets on the NEA database with radii up to 2.5 $R_{Jup}$ (open grey circles) and the seven exoplanet candidates identified in this work (black circles). The \textit{top} figure shows the orbital radii calculated using the orbital period and stellar mass in Table \href{https://zenodo.org/records/14618891?token=eyJhbGciOiJIUzUxMiJ9.eyJpZCI6ImE0YmE1ZTMyLTUyNDUtNDZhYi05NmI1LTUzYWM4NTgzYTVkYiIsImRhdGEiOnt9LCJyYW5kb20iOiIzNWFmZDRhMjExNjNlM2RjMjI4YzMyOGIyMTVmZjFjMSJ9.Q2GpAqZLxefIloSYfaJyUSOC8IenSx6kd6WeCiexA4oxUoyA6pfIg5TH16bcK2fUa_hCqroeC1LLBn8wSbaYzw}{Appendix A: Table 5}, i.e. Col. "Orb. Rad.(1)" in \href{https://zenodo.org/records/14618891?token=eyJhbGciOiJIUzUxMiJ9.eyJpZCI6ImE0YmE1ZTMyLTUyNDUtNDZhYi05NmI1LTUzYWM4NTgzYTVkYiIsImRhdGEiOnt9LCJyYW5kb20iOiIzNWFmZDRhMjExNjNlM2RjMjI4YzMyOGIyMTVmZjFjMSJ9.Q2GpAqZLxefIloSYfaJyUSOC8IenSx6kd6WeCiexA4oxUoyA6pfIg5TH16bcK2fUa_hCqroeC1LLBn8wSbaYzw}{Appendix A: Table 6} and the \textit{bottom} figure shows those calculated using the planet orbit-star radius ratio and the stellar radius in \href{https://zenodo.org/records/14618891?token=eyJhbGciOiJIUzUxMiJ9.eyJpZCI6ImE0YmE1ZTMyLTUyNDUtNDZhYi05NmI1LTUzYWM4NTgzYTVkYiIsImRhdGEiOnt9LCJyYW5kb20iOiIzNWFmZDRhMjExNjNlM2RjMjI4YzMyOGIyMTVmZjFjMSJ9.Q2GpAqZLxefIloSYfaJyUSOC8IenSx6kd6WeCiexA4oxUoyA6pfIg5TH16bcK2fUa_hCqroeC1LLBn8wSbaYzw}{Appendix A: Table 5}, i.e. Col. "Orb. Rad. (2)" in \href{https://zenodo.org/records/14618891?token=eyJhbGciOiJIUzUxMiJ9.eyJpZCI6ImE0YmE1ZTMyLTUyNDUtNDZhYi05NmI1LTUzYWM4NTgzYTVkYiIsImRhdGEiOnt9LCJyYW5kb20iOiIzNWFmZDRhMjExNjNlM2RjMjI4YzMyOGIyMTVmZjFjMSJ9.Q2GpAqZLxefIloSYfaJyUSOC8IenSx6kd6WeCiexA4oxUoyA6pfIg5TH16bcK2fUa_hCqroeC1LLBn8wSbaYzw}{Appendix A: Table 6}. The error bars reflect the low and high values shown in the table.} 
\label{fig:RvsA_chart}
\end{figure}

The exoplanet candidates described in this work have been compared with the confirmed exoplanets listed in the Nasa Exoplanet Archive\footnote{https://exoplanetarchive.ipac.caltech.edu/}. Fig. \ref{fig:RvsA_chart} shows the exoplanet radii plotted as a function of orbital radii. All the candidates lie within the clump of confirmed planets with radii between approximately 1 and 2 $R_{Jup}$ and with orbital radii between 0.01 and 0.1 A.U.

\begin{figure}
\centering
\includegraphics[width=\columnwidth]{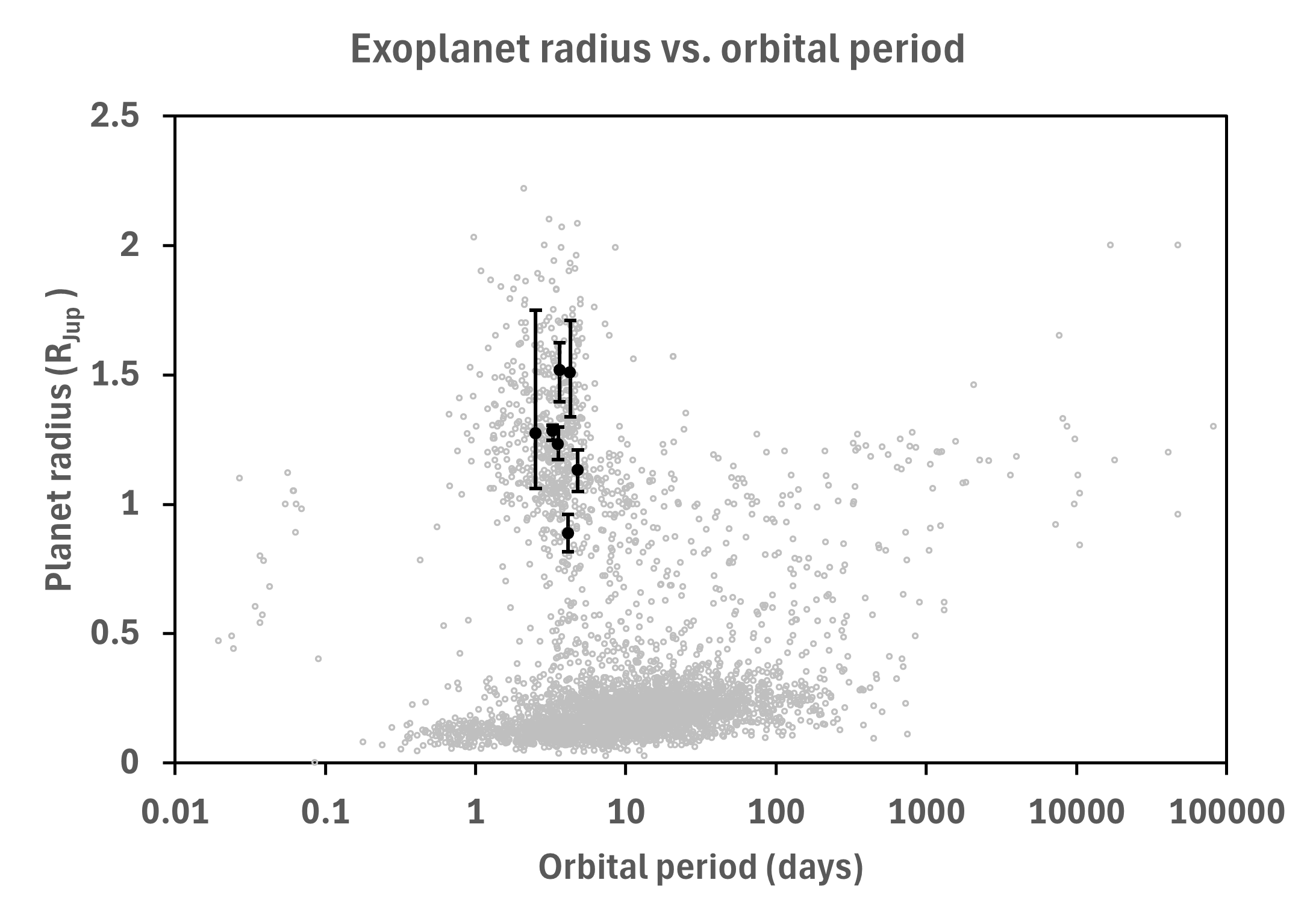}
\caption{This figure shows the confirmed exoplanets on the NEA database with radii up to 2.5 $R_{Jup}$ (open grey circles) and the seven exoplanet candidates identified in this work (black circles). The orbital periods are those shown in \href{https://zenodo.org/records/14618891?token=eyJhbGciOiJIUzUxMiJ9.eyJpZCI6ImE0YmE1ZTMyLTUyNDUtNDZhYi05NmI1LTUzYWM4NTgzYTVkYiIsImRhdGEiOnt9LCJyYW5kb20iOiIzNWFmZDRhMjExNjNlM2RjMjI4YzMyOGIyMTVmZjFjMSJ9.Q2GpAqZLxefIloSYfaJyUSOC8IenSx6kd6WeCiexA4oxUoyA6pfIg5TH16bcK2fUa_hCqroeC1LLBn8wSbaYzw}{Appendix A: Table 6}. The error bars reflect the low and high values shown in the table.} 
\label{fig:RvsP_chart}
\end{figure}

\begin{figure}
\centering
\includegraphics[width=\columnwidth]{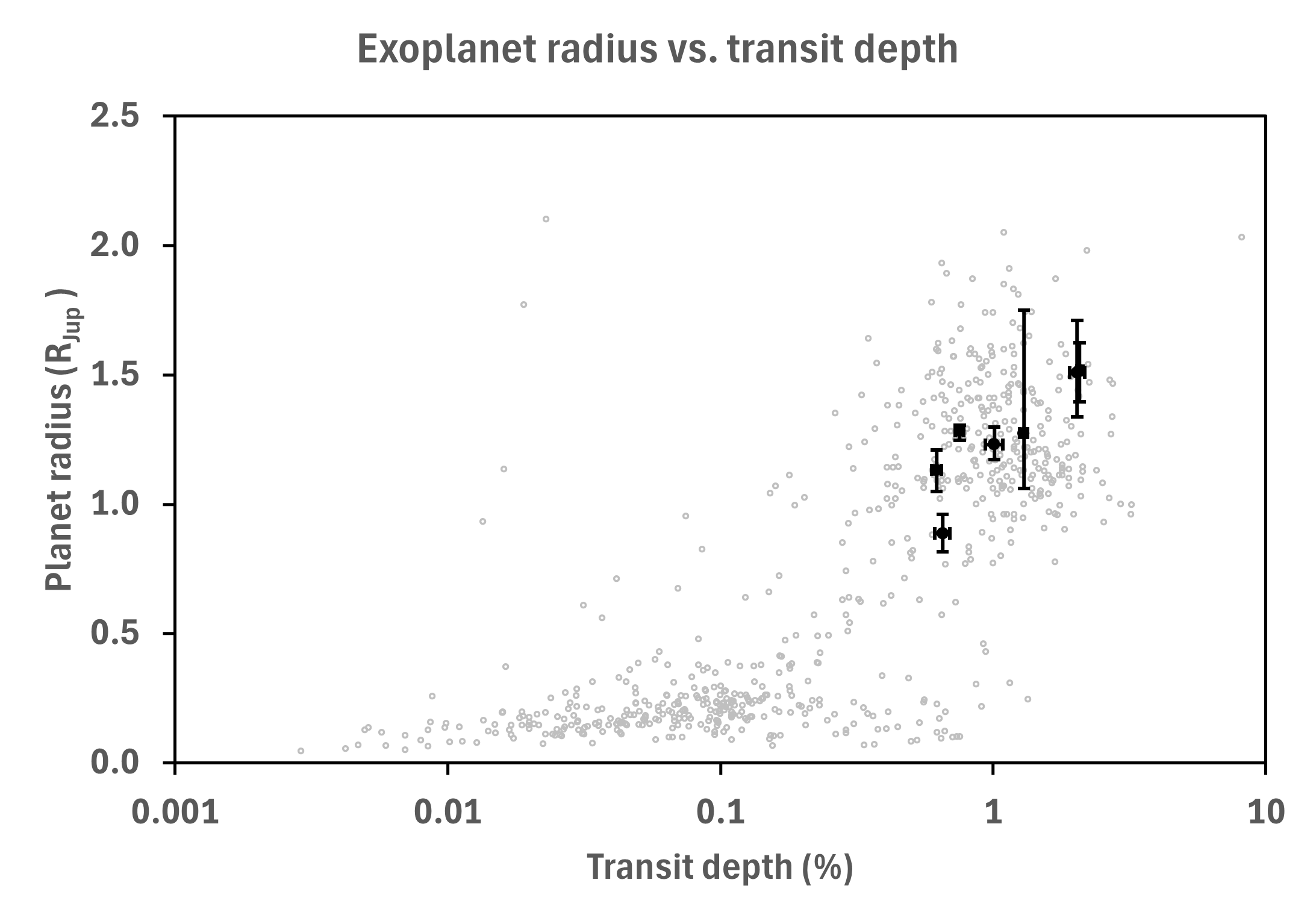}
\caption{This figure shows the confirmed exoplanets on the NEA database with radii up to 2.5 $R_{Jup}$ (open grey circles) and the seven exoplanet candidates identified in this work (black circles). The transit depths are taken from the minimum flux values derived from the Pytransit model for each candidate. The error bars reflect the $1\sigma$ uncertainty in the transit depth.} 
\label{fig:RvsTD_chart}
\end{figure}

Similarly, Figs. \ref{fig:RvsP_chart} and \ref{fig:RvsTD_chart} show the exoplanet radii plotted as functions of the orbital period and transit depth respectively. 

There is no generally accepted definition of a hot Jupiter, but \cite{2013ApJ...766...81F} define it as having $0.8 \le P \le 10.0$ days and $6.0 R_\oplus \le R_P \le 22.0 R_\oplus$ and \cite{2012ApJS..201...15H} define it as having $P \le 10.0$ days and $8.0 R_\oplus \le R_P \le 32.0 R_\oplus$. If we accept these definitions, then the candidates are all hot Jupiters.

\subsection{Selection bias} \label{sub:select_bias}

The exoplanet candidates described in this work all conform to the definition of a hot Jupiter. In this section we consider whether this narrow range may be due to selection bias.

\subsubsection{Planet radius bias} \label{subsub:planet_rad_bias}

The maximum radius limit for an exoplanet detection is determined by the arbitrary transit depth limit of 2.5\% described in Sect. \ref{sub:cand_sel_proc}. This equates to a planet of radius approximately 1.6 $R_{Jup}$, assuming a solar-sized host star. However, this does not allow for correction for light contamination or for larger sized host stars. Assuming a typical light contamination level of 20\% and a maximum host star radius of 1.5 $R_{\odot}$, then the largest detectable exoplanet will have a radius of approximately 2.6 $R_{Jup}$.

The minimum radius limit for an exoplanet detection is determined principally by the quality of the individual pixel light curves: from Equation \ref{equ:lc_search} it will immediately be seen that the lower the level of scatter in the light curve, the smaller the transit depth that can be detected. The stars examined in this work are generally fainter than the main target stars: the host stars for the eight candidates have an average $m_G$ of 15.96, about 2 magnitudes dimmer than the average main target. Moreover, the host star locations mostly lie outside of the main target apertures, and only the edges of their PSFs are detected. As a result, the light curves are generally weak and noisy in comparison with the main targets, and the possibility of detecting small planets is remote.

\begin{figure}
\centering
\includegraphics[width=\columnwidth]{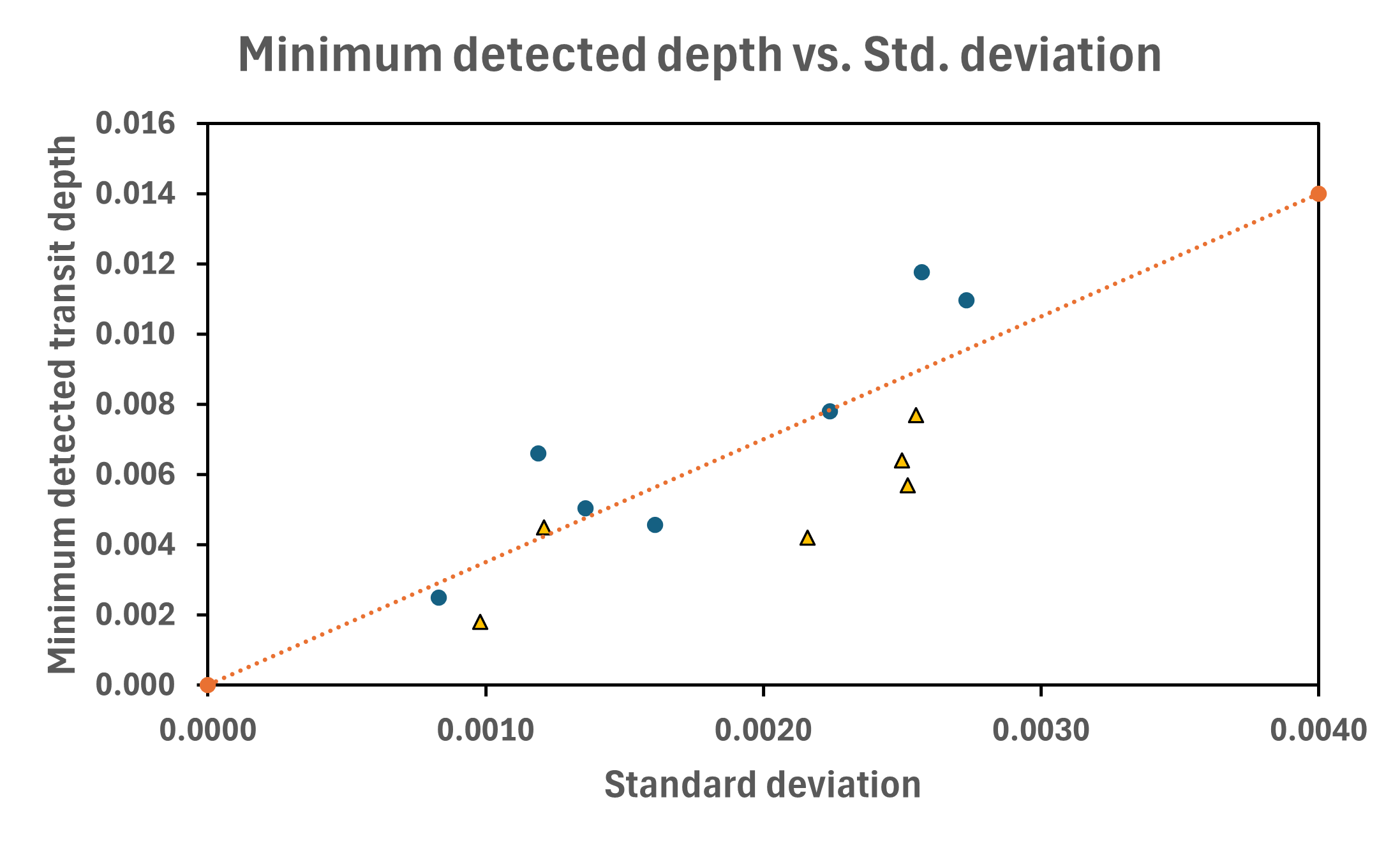}
\caption{This figure shows the minimum detectable fractional transit depth for the exoplanet candidates as a function of the standard deviation, $\sigma$, of the normalised pixel light curves (blue circles) and the normalised Q2-Q16 light curves (orange triangles). The red line is the expected lower limit of $3.5\sigma$.}
\label{fig:detect_limit}
\end{figure}

A realistic test for the smallest detectable planet was made by using the individual Q4 pixel light curves from which the exoplanet candidates were originally identified. For each light curve, the transit depths were progressively reduced until the search routine described in Sect. \ref{sub:search_proc} could no longer detect the transits. Fig. \ref{fig:detect_limit} shows the results of this, with the minimum detectable fractional transit depth plotted as a function of the standard deviation, $\sigma$, of the normalised light curve (blue circles). The red line is the expected lower detection limit of $3.5\sigma$. The minimum detection levels are mostly close to this line. The lowest detectable transit depth is 0.0025 for Gaia DR3 2100980735718445312 (KIC 4459924). The corresponding minimum and maximum planet sizes, of course, depend on the stellar radius, and Table \ref{tab:min_max_rad} shows the minimum and maximum planet radii for a range of stellar radii from 0.1 to 1.5 $R_{\odot}$. These are based on an upper limit of 0.025 fractional transit depth and a lower limit of 0.0025 fractional transit depth and include a correction for a typical light contamination level of 20\%.

A similar calculation was carried out using the full Q2-Q16 light curves. The minimum detectable fractional transit depth is 0.0018 for Gaia DR3 2100325529865507712 (KIC 4543171, resulting in slightly smaller minimum detectable planet sizes (shown in the right-hand column of Table \ref{tab:min_max_rad}).  The results are also shown in Fig. \ref{fig:detect_limit} (orange triangles). Note that one candidate was not detected at all in the Q2-Q16 light curves: Gaia DR3 2085335681690420608 (KIC 9674230). This arises from the merging of the Q4 (and Q8, Q12, Q16) light curves with the Q3, Q7, Q11 and Q15 light curves, which are of relatively poor quality, thus increasing the level of scatter in the Q2-Q16 lightcurve.

\begin{figure}
\centering
\includegraphics[width=\columnwidth]{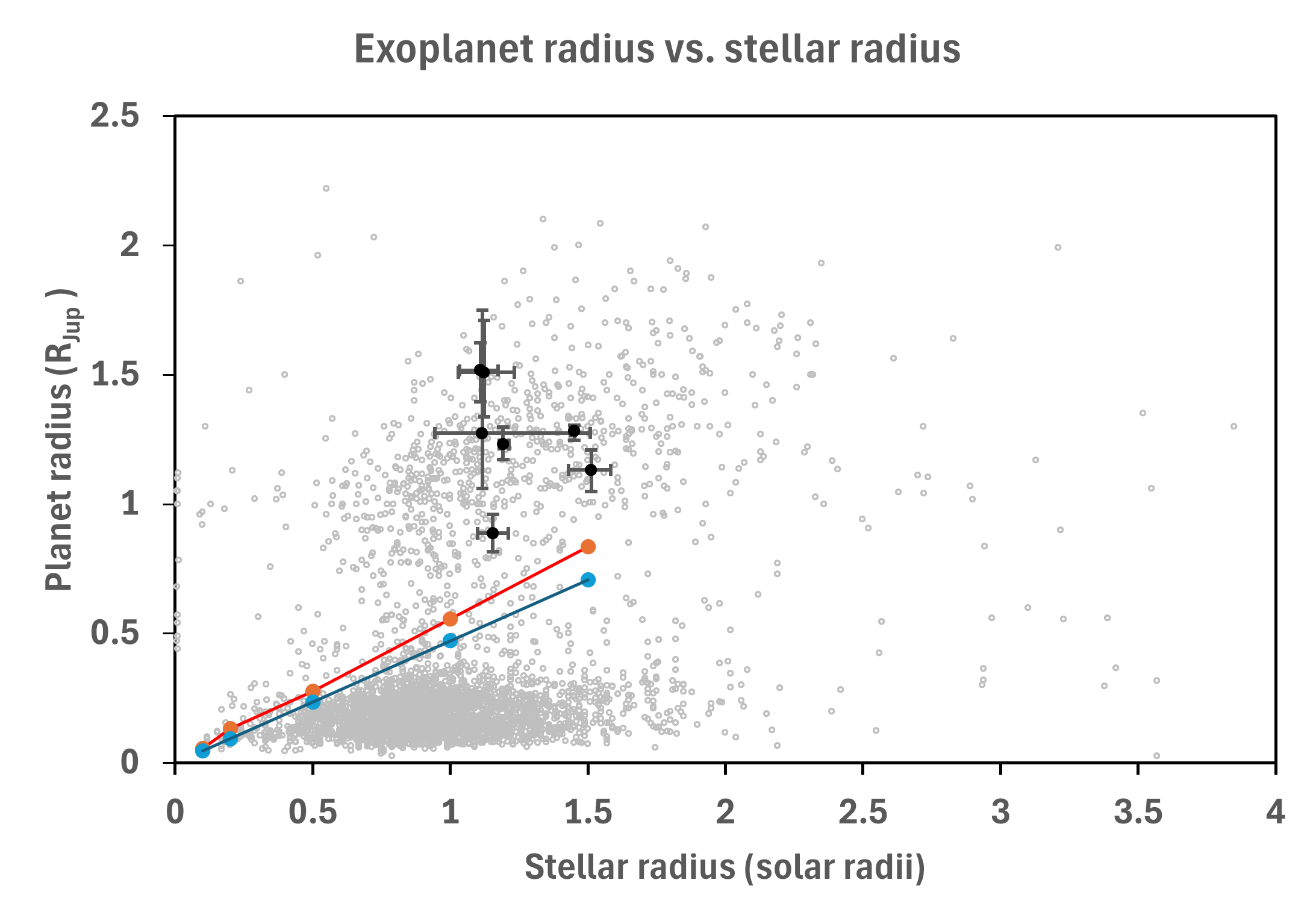}
\caption{This figure shows the confirmed exoplanets on the NEA database with radii up to 2.5 $R_{Jup}$ (open grey circles) and the seven exoplanet candidates identified in this work (black circles) plotted against their host star radii. The minimum detectable planet radii derived from the Q4 light curves (red circles and line) and from the Q2-Q16 light curves (blue circles and line) given in Table \ref{tab:min_max_rad} are also shown.} 
\label{fig:RvsRstar_chart}
\end{figure}

In principle then, earth-sized planets orbiting small stars should be detectable. However, Fig. \ref{fig:RvsRstar_chart} shows the radii of the known exoplanets listed in the NEA database plotted as a function of their host star radii, along with the exoplanet candidates. The red line on the chart shows the Q4 minimum detectable planet radii given in Table \ref{tab:min_max_rad} and the blue line shows the Q2-Q16 minimum detectable planet radii given in Table \ref{tab:min_max_rad}. These clearly indicate that planets with radius $< 0.5R_{Jup}$ are almost entirely outside the detectable range.

\begin{table}[ht]
\centering
\caption{Minimum and maximum detectable planet radii.
}
\begin{tabular*}{\columnwidth}{ c c c c }
\hline\hline
Stellar radius & Q4 Min. & Q4 Max. & Q2-Q16 Min. \\
 ($R_{\odot}$) & radius ($R_{Jup}$) & radius ($R_{Jup}$) & radius ($R_{Jup}$)\\
\hline
0.1 & 0.056 & 0.176 & 0.047 \\
0.2 & 0.111 & 0.352 & 0.094 \\
0.5 & 0.278 & 0.880 & 0.236 \\
1.0 & 0.557 & 1.761 & 0.472 \\
1.5 & 0.835 & 2.641 & 0.709 \\
\hline
\end{tabular*}
\label{tab:min_max_rad}
\end{table}

\subsubsection{Orbital period bias} \label{subsub:orbital_period_bias}

The search routine employed requires a minimum of 3 transits to be detected within the Q4 light curves. Thus, in principle, all exoplanets with periods up to 30 days should be detected (assuming a light curve of sufficiently good quality) and up to 45 days if the transits are fortuitously placed within the light curve. Clearly, this has not happened, with the maximum detected period being 4.791774 days for Gaia DR3 2100980735718445312 (KIC 4459924). However, Fig. \ref{fig:RvsP_chart} shows that known Jupiter-sized planets are concentrated in the period range $1 \le P \le 10$ days and comparatively few have longer periods. The chances of detecting one are low.

\section{Results} \label{sec:results1}

The results of the PyTransit modeling are shown in \href{https://zenodo.org/records/14618891?token=eyJhbGciOiJIUzUxMiJ9.eyJpZCI6ImE0YmE1ZTMyLTUyNDUtNDZhYi05NmI1LTUzYWM4NTgzYTVkYiIsImRhdGEiOnt9LCJyYW5kb20iOiIzNWFmZDRhMjExNjNlM2RjMjI4YzMyOGIyMTVmZjFjMSJ9.Q2GpAqZLxefIloSYfaJyUSOC8IenSx6kd6WeCiexA4oxUoyA6pfIg5TH16bcK2fUa_hCqroeC1LLBn8wSbaYzw}{Appendix A: Table 2}. These, combined with the stellar parameters shown in \href{https://zenodo.org/records/14618891?token=eyJhbGciOiJIUzUxMiJ9.eyJpZCI6ImE0YmE1ZTMyLTUyNDUtNDZhYi05NmI1LTUzYWM4NTgzYTVkYiIsImRhdGEiOnt9LCJyYW5kb20iOiIzNWFmZDRhMjExNjNlM2RjMjI4YzMyOGIyMTVmZjFjMSJ9.Q2GpAqZLxefIloSYfaJyUSOC8IenSx6kd6WeCiexA4oxUoyA6pfIg5TH16bcK2fUa_hCqroeC1LLBn8wSbaYzw}{Appendix A: Table 5} have been used to calculate the planetary parameters shown in \href{https://zenodo.org/records/14618891?token=eyJhbGciOiJIUzUxMiJ9.eyJpZCI6ImE0YmE1ZTMyLTUyNDUtNDZhYi05NmI1LTUzYWM4NTgzYTVkYiIsImRhdGEiOnt9LCJyYW5kb20iOiIzNWFmZDRhMjExNjNlM2RjMjI4YzMyOGIyMTVmZjFjMSJ9.Q2GpAqZLxefIloSYfaJyUSOC8IenSx6kd6WeCiexA4oxUoyA6pfIg5TH16bcK2fUa_hCqroeC1LLBn8wSbaYzw}{Appendix A: Table 6}.

These clearly indicate that all the candidates are hot Jupiter types, with planet radii ranging from 0.8878 to 1.5174 $R_{Jup}$ and orbital radii ranging from 0.0437 to 0.0653 A.U. or 0.0293 - 0.0495 A.U. depending on which calculations are used.

The light curves together with the fitted PyTransit models are shown in \href{https://zenodo.org/records/14618752?token=eyJhbGciOiJIUzUxMiJ9.eyJpZCI6ImY3OGNlYWNhLTBjZDQtNGIyYy1iNTJkLWFmYjk2NjJmZjRiNCIsImRhdGEiOnt9LCJyYW5kb20iOiIyMDVlNDcyNjJmMDQzMDMyOTllYzcwNDhjYWRiZTY5MSJ9.mhvK6uGYJ-4knrKz6fhhs-hvkdgU14SmEjqqxtKDG9uhognHf9vuCuarGdEaMAmj8f6USvVjDWkWU8Vqv1XplQ} {Appendix B}. The fits are good, with the phase-bin mean fluxes generally within one standard error of the model fit. The light curves all exhibit the characteristic shape of a full, rather than grazing, transit, thus supporting the planetary radii obtained from the Pytransit modeling.

Transit Timing Variation analysis has been applied to the light curves, but no evidence was found of second planets.

Comparison of the candidates with confirmed planets on the NEA database shows that the planetary parameters of the exoplanet candidates are all consistent with confirmed hot Jupiter exoplanets.

It should be noted that work by \cite{2012A&A...545A..76S} suggests a false positive rate of approximately 35\% for close-in Jupiter candidates, and it must be expected that only perhaps 4 or 5 of the candidates in this paper will eventually be confirmed.

\section{Summary} \label{sec:summary}

We have conducted a survey searching for pulsating stars and eclipsing binaries in the background pixels of the observed KIC targets in the original $Kepler$ field in the period range up to $\sim$ 90 days. In the course of this search, we have found seven exoplanet candidates which we have presented in this paper.

We prepared the light curves for each Quarter and each candidate for further analysis, that is, we detrended and normalised the Quarter light curves and then combined them to form $\sim$3.5 years long data sets (excluding Quarters 0, 1 and 17). We cross matched the pixel coordinates with the Gaia catalogue to identify the host star candidates.

We have analysed the transit light curves using the PyTransit software package and hence determined the planetary parameters. Using the stellar data available from the \textit{Gaia} catalogue, we have estimated the candidate planet radii and orbital radii. All of the seven candidates are consistent with hot Jupiter type planets. The planetary parameters are also consistent with the known exoplanets listed on the NEA database. 

We have carried out candidate vetting, including odd-even transit modelling and centroid analysis to strengthen the case for the candidates being exoplanets.

We have carried out Transit Timing Variation analysis to search for second planets in the systems, but no evidence for these was found.

The candidate light curves taken together with the Pytransit models and the comparison with known exoplanets provide strong support for these candidates being exoplanets.

The results of the survey for pulsating variables and long period eclipsing binaries will also be presented in subsequent papers.

\section{Data Availability} \label{sec:data_availabilty}

\href{https://zenodo.org/records/14618891?token=eyJhbGciOiJIUzUxMiJ9.eyJpZCI6ImE0YmE1ZTMyLTUyNDUtNDZhYi05NmI1LTUzYWM4NTgzYTVkYiIsImRhdGEiOnt9LCJyYW5kb20iOiIzNWFmZDRhMjExNjNlM2RjMjI4YzMyOGIyMTVmZjFjMSJ9.Q2GpAqZLxefIloSYfaJyUSOC8IenSx6kd6WeCiexA4oxUoyA6pfIg5TH16bcK2fUa_hCqroeC1LLBn8wSbaYzw}{Appendix A: Tables of results}

\href{https://zenodo.org/records/14618752?token=eyJhbGciOiJIUzUxMiJ9.eyJpZCI6ImY3OGNlYWNhLTBjZDQtNGIyYy1iNTJkLWFmYjk2NjJmZjRiNCIsImRhdGEiOnt9LCJyYW5kb20iOiIyMDVlNDcyNjJmMDQzMDMyOTllYzcwNDhjYWRiZTY5MSJ9.mhvK6uGYJ-4knrKz6fhhs-hvkdgU14SmEjqqxtKDG9uhognHf9vuCuarGdEaMAmj8f6USvVjDWkWU8Vqv1XplQ}{Appendix B: Pytransit model fits}

\href{https://zenodo.org/records/14618766?token=eyJhbGciOiJIUzUxMiJ9.eyJpZCI6IjczMzliOGEyLWYyOGUtNDliZi1hYTM5LTk0Y2EwNjhjMmY1ZCIsImRhdGEiOnt9LCJyYW5kb20iOiI4YTg0YThlYTM1YzcyNDViYzU0ODdjNmVlMTQwNWI5ZiJ9.H63XLMFfgfqgfXsH2vTl-duISKPQShy3cOD4KE89q_os5GVzPFybxrTMgEGpARfGEKt5D5sJ-bo0GkGF1uGOAg}{Appendix C: PyTransit odd and even transit model fits}

\begin{acknowledgements} \label{sec:ack}

This paper includes data collected by the Kepler mission. Funding for the Kepler mission is provided by the NASA Science Mission directorate.

This research was supported by the KKP-137523 'SeismoLab’ \'Elvonal grant and the SNN-147362 grant of the Hungarian Research, Development and Innovation Office (NKFIH).

This work has made use of data from the European Space Agency (ESA) mission {\it Gaia} (\url{https://www.cosmos.esa.int/gaia}), processed by the {\it Gaia} Data Processing and Analysis Consortium (DPAC,
\url{https://www.cosmos.esa.int/web/gaia/dpac/consortium}). Funding for the DPAC has been provided by national institutions, in particular the institutions participating in the {\it Gaia} Multilateral Agreement.

This research has made use of the NASA Exoplanet Archive, which is operated by the California Institute of Technology, under contract with the National Aeronautics and Space Administration under the Exoplanet Exploration Program.

We have also made use of the VizieR catalogue access tool, CDS, Strasbourg, France (DOI: 10.26093/cds/vizier). The original description of the VizieR service was published in A\&AS 143, 23.

The research was also carried out with the use of a number of other facilities which we gratefully acknowledge:
\begin{enumerate}
    \item PyTransit, a Python package for the analysis and modelling of transiting exoplanet light curves \citep{Parviainen_2015}.
    \item Lightkurve, a Python package for Kepler and TESS data analysis \citep{2018ascl.soft12013L}.
    \item The "Aladin sky atlas" developed at CDS, Strasbourg Observatory, France \citep{2000A&AS..143...33B}.
    \item Astropy,\footnote{http://www.astropy.org} a community-developed core Python package for Astronomy \citep{astropy:2013}.
    \item PSFmachine, a Python package for deblending \textit{Kepler} light curves \citep{2023ascl.soft06056H}.
    \item Vetting, a Python package for centroid testing of transits and eclipses \citep{Hedges_2021}
\end{enumerate}
\end{acknowledgements}

\bibliography{exopaper} 

\end{document}